\documentclass{sig-alternate}

\usepackage{latexsym}
\usepackage{amsmath}
\usepackage{amsfonts}
\usepackage[english]{babel}
\usepackage{array}

\newtheorem{definition}{Definition}[section]
\newtheorem{theorem}{Theorem}

\def\punto{\hspace*{\fill}\Box}

\begin{document}
%
\conferenceinfo{CIKM'11,} {October 24--28, 2011, Glasgow, Scotland, UK.}
\CopyrightYear{2011}
\crdata{978-1-4503-0717-8/11/10}
\clubpenalty=10000
\widowpenalty = 10000

\title{Effective Retrieval of Resources in Folksonomies Using a New Tag Similarity Measure}

%
%
%
%
%
\numberofauthors{3}

\author{
%
%
\alignauthor
Giovanni Quattrone and Licia Capra\\
	\affaddr{Dept. of Computer Science}\\
	\affaddr{University College London, UK}\\
	\email{\footnotesize \{g.quattrone,l.capra\}@cs.ucl.ac.uk}
\alignauthor
Pasquale De Meo and Emilio Ferrara\\
    \affaddr{Dept. of Physics, Dept. of Mathematics}\\
    \affaddr{University of Messina, Italy}\\
    \email{\footnotesize \{pdemeo,eferrara\}@unime.it}	
\alignauthor
Domenico Ursino\\
	\affaddr{DIMET}\\
	\affaddr{Universit\`a Mediterranea di Reggio Calabria, Italy}\\
	\email{\footnotesize ursino@unirc.it}
}

\date{}

\maketitle

\begin{abstract}
Social (or folksonomic) tagging has become a very popular way to describe content within Web 2.0 websites. However, as tags are informally defined, continually changing, and ungoverned, it has often been criticised for lowering, rather than increasing, the efficiency of searching. To address this issue, a variety of approaches have been proposed that recommend users what tags to use, both  when labeling and when looking for resources. These techniques work well in dense folksonomies, but they fail to do so when tag usage exhibits a power law distribution, as it often happens in real-life folksonomies. To tackle this issue, we propose an approach that {\em induces} the creation of a dense folksonomy, in a fully automatic and transparent way: when users {\em label} resources, an innovative tag similarity metric is deployed, so to enrich the chosen tag set with related tags already present in the folksonomy.  The proposed metric, which represents the core of our approach, is based on the mutual reinforcement principle. Our experimental evaluation proves that the accuracy and  coverage of searches guaranteed by our metric are higher than those achieved by applying classical metrics.
\end{abstract}




\vfill \eject \section{Introduction}
\label{sec:introduction}

Social media applications, such as blogs, multimedia sharing sites, question and answering systems, wikis and online forums, are growing at an unprecedented rate and are estimated to generate a significant amount of the content currently available on the Web. This has exponentially increased the amount of information that is available to users, from videos on sites like YouTube and MySpace, to pictures on Flickr, music on Last.fm, blogs on Blogger, and so on. This content is no longer categorised according to pre-defined taxonomies (or ontologies). Rather, a new trend called {\em social} (or {\em folksonomic}) {\em tagging} has emerged, and  \linebreak quickly become the most popular way to describe content within Web 2.0 websites. Unlike taxonomies, which overimpose a hierarchical categorisation of content, folksonomies empower end users by enabling them to freely create and choose the tags that best describe a piece of information (a picture, a document, a blog entry, a video clip, etc.). However, this freedom comes at a cost: since tags are informally defined, continually changing, and ungoverned, finding content of interest has become a main challenge because of the number of synonyms, homonyms and polysemies, as well as the inevitable heterogeneity of users and the noise they introduce.

In order to assist users finding content of their own interest within this information abundance, new approaches, inspired by traditional recommender systems, have been developed \cite{FiNePa07,ZaCa08,TsMaSc08}. These  often exploit an underlying  tag similarity measure; whenever a user labels a resource or searches for it by adopting a set of tags, they suggest new tags to be added to the resource label or to the user query, on the basis of their {\em similarity} to the original tags expressed by the user herself. They do so to increase the chances of finding content of relevance in these extremely sparse settings.

Various classic metrics have been used to compute tag similarity, including, for instance, cosine similarity, Jaccard coefficient, and Pearson Correlation. Some of the approaches exploiting these metrics \cite{jaschke2007tag,gemmell*09} have proved to  achieve excellent results; however, they do so {\em only if} the underlying folksonomy is {\em already dense}, and they operate by making it even denser. Nevertheless, we observe that this assumption does not hold true; rather, most real life folksonomies exhibit a \emph{power law} distribution of tag usage \cite{cattuto2007network,Ursino-InfSys-09}, with few tags labeling most resources, and most tags labeling just a few resources instead. This means that, in practical cases, if we were select any two tags, the probability that the resources jointly labeled by them is non-zero is extremely low. As a result, computing tag similarity on real folksonomies, using traditional metrics like cosine similarity, would almost always yield close-to-zero values, thus failing to support users in retrieving resources relevant to their queries.

In this paper, we propose an approach that transparently {\em induces} the creation of a {\em dense folksonomy}, thus supporting the effective retrieval of resources {\em by construction}. At the core of our approach lies an innovative tag similarity metric, used to recommend tags both when labeling resources and when querying the folksonomy. This metric is based on the  {\em mutual reinforcement} principle, and thus computed following an iterative algorithm: two tags are deemed similar if they label similar resources, and vice-versa, two resources are similar if they have been labeled by similar tags. When a user labels a new resource,  or when she is submitting a query to retrieve some resources, the above metric is used to automatically expand the user-selected tag set with those tags, already present within the folksonomy, that are: {\em (i)} most similar to those she initially submitted, and {\em (ii)} among those most widely used in the folksonomy.

We have conducted an extensive experimental evaluation on two large-scale datasets, namely BibSonomy and CiteULike. The obtained results demonstrate that our similarity metric operates effectively even in very sparse settings, where traditional metrics (including  cosine similarity, SimRank \cite{JeWi02} and Latent Semantic Indexing (LSI) \cite{MaRaSc08}) fail.

\section{Description of our approach}
\label{sec:approach}

In this section we provide a detailed description of our approach to support effective resource retrieval in large-scale folksonomies. Before illustrating it, we formalize the concept of a folksonomy as done in \cite{hotho2006information}.

\begin{definition} \label{def:folk}
    Let $US = \{u_1, \ldots, u_{n_u} \}$ be a set of  users,  $RS = \{r_1,\ldots, r_{n_r}\}$ a set of resource URIs, and \linebreak $TS = \{t_1, \ldots,t_{n_t}\}$  a set of tags. A {\em folksonomy} $F$ is a tuple $F = \langle US,RS,TS, AS \rangle $, where $AS \subseteq US \times RS \times TS$ is a ternary relationship called {\em tag assignment set}. $\punto$
\end{definition}

In the above definition we do not make any assumption about the nature of resources; they could be a URL associated with a Web page (like in Delicious), photos (as in Flickr), music files (as in Last.fm), documents (as in CiteULike), and so on.

According to Definition~\ref{def:folk}, a folksonomy $F$ is a ``three-dimensional'' data structure whose ``dimensions'' are represented by users, tags and resources. In particular, an element $a \in AS$ is a triple $\langle u, r, t \rangle$, indicating that  user $u$ labeled  resource $r$ with  tag $t$. To simplify folksonomy modelling and management, the inherent tripartite graph structure is often mapped into three  matrices, whereby each matrix models one relationship at a time \cite{mika2007ontologies}.

In this paper, we adopt the same matrix-based representation. Specifically, the association between tags and resources can be modelled by a $n_t \times n_r$ matrix $\mathbf{TR}$, called \textsl{Tag-Resource} matrix, being $n_t$ and $n_r$ the number of tags and resources, respectively. The generic entry of such a matrix $\mathbf{TR}_{ij}$ is the number of times the $i^{th}$ tag labels the $j^{th}$ resource. In an analogous fashion we can introduce the \textsl{Tag-User} and the \textsl{Resource-User} matrices, $\mathbf{TU}$ and $\mathbf{RU}$.

Our approach consists of two phases: the former, executed offline,  computes pairwise tag similarities,  by means of an innovative tag similarity metric. The second phase, executed in real-time (online),  follows a  well-established process: when a user is labeling a new resource, or is querying the system to retrieve some resources, the tag set she has chosen is automatically expanded using the tags that are deemed most related to the user-elected ones, based on the similarities previously computed. We now illustrate each phase in more detail; we then conclude this section with a discussion on how our approach can be efficiently realised in practice.

\subsection{Phase 1: Tag Similarity Computation}
\label{sub:Phase1}

As previously pointed out, this phase aims at computing pairwise similarity of all tags in use within the folksonomy. A variety of metrics have been proposed in the literature, mostly based on tag co-occurrence (such as cosine similarity); however, we claim these approaches fail to work in the sparse settings we target. To see why, let us consider cosine similarity. Given an arbitrary pair of tags $t_i$ and $t_j$, their cosine similarity $s(t_i,t_j)$ would be computed as
    \begin{equation}
    \label{eqn:ressim}
          s(t_i,t_j) = \frac{\langle \mathbf{t}_{r}(i) , \mathbf{t}_{r}(j)\rangle}{\sqrt{\langle \mathbf{t}_{r}(i) , \mathbf{t}_{r}(i)\rangle} \cdot \sqrt{\langle \mathbf{t}_{r}(j) , \mathbf{t}_{r}(j)\rangle}}
    \end{equation}
where $\mathbf{t}_{r}(i)$ and $\mathbf{t}_{r}(j)$ denote the $i^{th}$ and the $j^{th}$ row of $\mathbf{TR}$.

Equation~\ref{eqn:ressim} states that the similarity score of a pair of tags is high if they {\em jointly co-occur} in labeling the same subset of resources. One important underlying assumption must hold for cosine similarity to work well: matrix $\mathbf{TR}$ must be densely populated. Unfortunately, this assumption does not hold in real folksonomies.

As an example, let us consider a real-world folksonomy like BibSonomy. BibSonomy \cite{hotho2006bibsonomy,jaschke2007analysis} is a social bookmarking service in which users are allowed to tag both URLs and scientific papers. A {\em power law} distribution of tags on scientific references emerges. In particular, roughly 81\% of resources were described by no more than 5 different tags (and roughly 58\% by less than 3 ). Furthermore, there is a small portion of frequently adopted tags, and a long tail of tags (roughly 81\%) being used less than 5 times overall. Matrix $\mathbf{TR}$ is thus rather sparse: if we were to select any pair of tags $t_i$ and $t_j$, most of the components of the corresponding vectors $\mathbf{t}_{r}(i)$ and $\mathbf{t}_{r}(j)$ would be 0, and so would be their inner product. In other words, the cosine similarity between {\em any} two folksonomy tags would be very close to 0, regardless of what the selected tags are; recommending tags (in Phase 2) based on such metric would thus be unfruitful.

Although classical similarity measures based on \linebreak co-occurrences are inadequate in scenarios characterized by power law distributions, other metrics have been proposed and successfully used in Information Retrieval which could potentially be applied in our domain. We considered, in particular,  one of the state-of-the-art techniques, namely \emph{Latent Semantic Indexing} (LSI) \cite{MaRaSc08}. LSI approximates the matrix $\mathbf{TR}$ by computing its \emph{top} $k$ eigenvalues. This is equivalent to mapping $\mathbf{TR}$ onto a low-dimensional vectorial space, with $k$ dimensions (called \emph{Latent Space}). Similarities between tags (respectively, resources) are computed in the Latent Space by applying the cosine similarity. Unfortunately, the application of this technique raised several concerns, because: {\em (i)} the computation of LSI on large matrices is very costly (and could indeed be practically unfeasible in real folksonomies); {\em (ii)} the tuning of  parameter $k$ is complex and time-expensive, and the quality of the produced results is very sensitive to such  value.

More suitable to the folksonomy domain are techniques that rely on the {\em mutual reinforcement principle}. One of the most popular techniques based on it is {\em SimRank} \cite{JeWi02}. SimRank uses an \emph{iterative approach} to compute similarities whereby, in each iteration, the similarity between any two objects (be them tags or resources) is computed, based on the similarities already computed in the previous iteration.

If we were to adopt SimRank, the equations used at the $k^{th}$ iteration would be
    \begin{eqnarray}
    	{st}^{k}(t_a,t_b) = \frac{C_1}{|r(t_a)| \cdot |r(t_b)|} \sum_{r_i \in r(t_a)} \sum_{r_j \in r(t_b)} sr^{k-1} (r_i,r_j) && \label{eqn:simranktag}\\
    	{sr}^{k}(r_a,r_b) = \frac{C_2}{|t(r_a)| \cdot |t(r_b)|} \sum_{t_i \in t(r_a)} \sum_{t_j \in t(r_b)} st^{k-1} (t_i,t_j) && \label{eqn:simrankres}
    \end{eqnarray}
where: {\em (i)} ${st}^k(t_a, t_b)$ (resp, ${sr}^k(r_a, r_b)$) denotes the similarity between $t_a$ and $t_b$ (resp, $r_a$ and $r_b$) at the $k^{th}$ iteration; {\em (ii)} the set $r(t_a)$ (resp., $t(r_a)$) is the set of resources (resp, tags) associated with $t_a$ (resp., $r_a$); {\em (iii)} $C_1$ and $C_2$ are two {\em normalization constants} belonging to the real interval $[0,1]$.

Equations~\ref{eqn:simranktag}--\ref{eqn:simrankres} suffer from some main drawbacks that limit their applicability in our setting:

\begin{itemize}

\item SimRank does not take into account the number of times a  tag intervenes in labeling a resource, thus discarding valuable information available within the folksonomy.

\item SimRank does not distinguish between tags that have labeled exactly the same resource, and tags that (by means of one or more iterative steps) have been labeling related, but different, resources.

\end{itemize}

Although we share with SimRank the idea of computing tag/resource similarity by means of an application of the mutual reinforcement principle, we advocate for some main changes, in line with the discussion above. To begin with, the frequency with which a tag intervenes in labeling a resource is a very important piece of information that should be leveraged in the similarity computation process. Furthermore, a new factor (which we  will call {\em mutual reinforcement factor}) should be introduced,  to give more relevance to tags that labeled the very {\em same} resources, with respect to those that labeled related (but not the very same) resources.

We have thus derived a novel similarity metric,  specifically conceived for the folksonomy setting. In detail, the similarity computation is performed recursively. For the base case, given a pair of tags $\langle t_a, t_b \rangle$ and a pair of resources $\langle r_a, r_b \rangle$, the {\em tag similarity} ${st}^0(t_a,t_b)$ and the {\em resource similarity} ${sr}^0(r_a,r_b)$ is defined as follows
    \begin{equation}
    	\label{eqn:basecase}
    	{st}^0(t_a,t_b) = \delta_{ab} \quad \quad {sr}^0(r_a,r_b) = \delta_{ab}
    \end{equation}

Equation~\ref{eqn:basecase} states that, in the initial step, each tag (resp., resource) is similar only to itself and it is dissimilar to all other tags (resp., resources).

At the $k^{th}$ step, let ${st}^{k-1}(t_a,t_b)$ (resp., ${sr}^{k-1}(r_a,r_b)$) be the tag (resp., resource) similarity between $t_a$ and $t_b$ (resp., $r_a$ and $r_b$). The following rules can be applied to compute ${st}^k(t_a,t_b)$ (resp., ${sr}^k(r_a,r_b)$)
    \begin{eqnarray}
    	{st}^{k}(t_a,t_b) = \frac{ ST^k(t_a,t_b) }{ \sqrt{ST^k(t_a,t_a)} \cdot \sqrt{ST^k(t_b,t_b)} } && \label{eqn:coSim1}\\
    	{sr}^{k}(r_a,r_b) = \frac{ SR^k(r_a,r_b) }{ \sqrt{SR^k(r_a,r_a)} \cdot \sqrt{SR^k(r_b,r_b)} } && \label{eqn:coSim2}
    \end{eqnarray}
where
    \begin{eqnarray}
        ST^{k}(t_a,t_b) = \sum_{i,j=1}^{n_r}{ \mathbf{TR}_{ai} \cdot \Psi_{ij} \cdot {sr}^{k-1}(r_i,r_j) \cdot \mathbf{TR}_{bj} } && \label{eqn:defSTSR1} \\
        SR^{k}(r_a,r_b) = \sum_{i,j=1}^{n_t}{ \mathbf{TR}_{ia} \cdot \Psi_{ij} \cdot {st}^{k-1}(t_i,t_j) \cdot \mathbf{TR}_{jb} } && \label{eqn:defSTSR2}
    \end{eqnarray}

Here $\Psi_{ij}$ is equal to $1$ if $i = j$, while  it is equal to $\psi$ if $i \neq j$.  $\psi$ is what we call {\em mutual reinforcement factor}, and is a value belonging to the real interval $[0,1]$.

Equations~\ref{eqn:coSim1}--\ref{eqn:coSim2} rely on the following intuitions. Given a pair of tags $\langle t_a, t_b \rangle$, at the $k$ iteration, we consider {\em all  pairs of resources} $\langle r_i, r_j\rangle$ in the folksonomy and we take their similarity ${sr}^{k-1}(r_i,r_j)$ into account to compute ${st}^{k}(t_a,t_b)$. In particular, we compute a {\em weighted sum} of all the similarity values ${sr}^{k-1}(r_i,r_j)$, where the weights reflect the {\em strength of the association} between the tag $t_a$ and the resource $r_i$, and the tag $t_b$ and the resource $r_j$. As a consequence, the higher the similarity between $r_i$ and $r_j$, the higher the contribution of the association between $t_a$ and $r_i$, as well as $t_b$ and $r_j$. Finally, the mutual reinforcement factor $\psi$ is instrumental to give higher relevance to tags that labeled the very {\em same} resources, (resp., to resources labeled by the very same tags): indeed, the higher $\psi$, the higher the relevance assigned to  similar  resources (resp., tags) in the tag (resp., resource) similarity computation.

We argue that Equations~\ref{eqn:coSim1}--\ref{eqn:coSim2} are able to effectively address the power law challenge we outlined above. In fact, when computing tag (resp.,  resource) similarity, our measure leverages  the similarity of all pairs of resources (resp., tags) in the folksonomy. While cosine similarity restricts its attention to those resources jointly labeled by two tags  (resp., those tags jointly labeling two resources), which are usually very few, our metric  {\em iteratively propagates}  similarity scores by considering all the pairs of similar resources jointly labeled by the two tags    (resp., all the pairs of similar tags jointly labeling two resources). In this way, our measure can be applied in settings characterized by power law distributions of tag usage.

Similarly to state-of-the-practice recommender systems, we expect the above tag similarity computation to be performed offline, at regular intervals of time (e.g., daily, \linebreak weekly), depending on the growth of the system and the available computational resources. With these similarities \linebreak pre-computed, we now proceed to discuss how tag recommendation and expansion is performed, both when labeling new resources and when querying the folksonomy.

\subsection{Phase 2: Tag Expansion}
\label{sub:Phase2}

Key to our approach is the use of the previously computed tag similarities to automatically  expand the tag set chosen by the user, both when labeling a new resource and/or when querying the folksonomy. Note that, by performing tag expansion upon {\em adding} a new resource in the system, we implicitly {\em induce the creation of a denser folksonomy}; furthermore, by performing tag expansion upon {\em querying} the system using the very same approach, we implicitly {\em induce the community to use a common vocabulary}. Taken together, these tag expansions have the effect of providing more accurate answers to users' searches within large-scale folksonomy.

The approach can be summarised as follow. Let  $tSet = \{ t_1, \ldots, t_n \}$ be the set of user-selected tags, either  to label a resource or to submit a query in the folksonomy. Let  $t_j$ be a tag in $tSet$ and $t_i$ a tag not in $tSet$. We assign a score $sc(t_i, t_j)$ to $t_i$ with respect to $t_j$ based on: {\em (i)} $st(t_i,t_j)$ - the similarity between $t_i$ and $t_j$ as previously computed; {\em (ii)} $count(t_i)$ - the number of times $t_i$ appears in the folksonomy; {\em (iii)} $IRF(t_i)$ - the inverse resource frequency of $t_i$ (similar to IDF in Information Retrieval). This is a measure of the general importance of $t_i$ within the whole folksonomy, and it is obtained by dividing the total number of resources in the folksonomy by the number of resources labeled by $t_i$, and then taking the logarithm of the quotient.

More precisely, $sc(t_i,t_j)$ is computed as follows
    \begin{equation}
    	\label{eqn:singleScore}
        sc(t_i, t_j) = st(t_i, t_j) \cdot \log{count(t_i)} \cdot IRF(t_i)
    \end{equation}

Equation~\ref{eqn:singleScore} assigns high scores to those tags that are both similar to $t_j \in tSet$ (as per our similarity metric) {\em and}, crucially, which are both {\em largely used} ($count(t_i)$) and {\em important} ($IRF(t_i)$) in the overall folksonomy. Intuitively, our approach  expands user-selected tags with related tags that are part of the emerging common vocabulary of widely used tags. Note that we compute the logarithm of $count(t_i)$ to give equal weight to frequently used tags and  to important ones (as computed by  $IRF(t_i)$, which, by definition, already computes the logarithm).

Finally, the total score $SC$ of $t_i$ with respect to $tSet$ is obtained by summing the scores of $t_i$ with respect to all the tags of $tSet$
    \begin{equation}
    	\label{eqn:Score}
        SC(t_i, tSet) = \sum_{t_j \in tSet}{sc(t_i, t_j)}
    \end{equation}

Although in this paper we will be evaluating a fully automatic approach, whereby  the user-selected set of tags $tSet$ is transparently expanded with the  $k$ highest scoring tags according to Equation~\ref{eqn:Score}, a more interactive approach could be adopted, whereby users are suggested up to $k$ expansion tags, and they can decide which ones, if any, to use. Such an approach may  lead to even more accurate results than those we will report in Section~\ref{sec:evaluation}, and it is thus worth exploring in the future,  by means of controlled user studies.

\subsection{Taming Computational Complexity}
\label{sub:cost}

The practical usability of our approach is strictly linked to the computational complexity of  Equations~\ref{eqn:coSim1}--\ref{eqn:coSim2}. In particular:

\begin{itemize}

\item From a theoretical standpoint, the computation of each pairwise tag similarity may require an infinite number of iterations. As a consequence, a stopping criterium is required so that the execution of Equations~\ref{eqn:coSim1}--\ref{eqn:coSim2} terminates after a finite (and hopefully low) number of iterations.

\item Equation~\ref{eqn:coSim1} (resp., Equation~\ref{eqn:coSim2}) requires the computation of $n_r^2$ resource-resource (resp,.  $n_t^2$ tag-tag) similarities, at each  $k^{th}$ step. This could make our similarity measure inapplicable in practical cases, because each iteration would require exactly $n_r^2 \times n_t^2$ computations.

\end{itemize}

However, we can prove that these theoretical limits do not apply in practice, and that, in fact, our new tag similarity metric requires a computational complexity comparable to that of cosine similarity. First of all, convergence of Equations~\ref{eqn:coSim1}--\ref{eqn:coSim2} has been demonstrated.

\begin{theorem}
\label{the:conv}
    Let ${st}^{k}(t_a,t_b)$ and ${sr}^{k}(r_a,r_b)$ be defined as in Equations~\ref{eqn:coSim1}--\ref{eqn:coSim2}. Given any pair of tags $t_a$ and $t_b$ and any pair of resources $r_a$ and $r_b$, the sequences ${st}^{k}(t_a,t_b)$ and ${sr}^{k}(r_a,r_b)$ converge. $\punto$
\end{theorem}

The proof of the above theorem is available in full  \linebreak at {\small \tt http://tinyurl.com/proof-cikm2011} and is based on the demonstration that the sequences ${st}^{k}(t_a,t_b)$ and ${sr}^{k}(r_a,r_b)$ are both bounded and not-decreasing. To complement this theoretical result, our experiments on two real folksonomies (Section~\ref{sub:scalability}) will provide evidence of very fast convergence indeed.

The second important result is that Equations~\ref{eqn:coSim1}--\ref{eqn:coSim2} can be defined, without any loss of generality, as simple matrix products (such as in cosine similarity). Specifically, let $\mathbf{st}^k$ and $\mathbf{sr}^k$ be the tag-tag and resource-resource similarity matrices, respectively, with $\mathbf{st}^0=\mathbf{I}_t$ and $\mathbf{sr}^0=\mathbf{I}_r$, where $\mathbf{I}_t$ (resp., $\mathbf{I}_r$) is the $n_t \times n_t$ (resp., $n_r \times n_r$) {\em identity matrix}. We use symbol ``$\circ$'' to refer to the  Hadamard matrix product \cite{GoVa96}. At the $k^{th}$ step the $\mathbf{st}^k$ and $\mathbf{sr}^k$ matrices are computed as
    \begin{equation}
    	\mathbf{{st}}^{k} = \mathbf{ST}^k \circ \mathbf{DT}^k \quad \mbox{and} \quad \mathbf{{sr}}^{k} = \mathbf{SR}^k \circ \mathbf{DR}^k  \label{eqn:matrices}
    \end{equation}
where
    \begin{eqnarray}
    	& \mathbf{ST}^{k} =\mathbf{TR} \times \left(\mathbf{\Psi_r} \circ \mathbf{{sr}}^{k-1}\right) \times \mathbf{TR}^t & \label{eqn:ST}\\
    	& \mathbf{SR}^{k} = \mathbf{TR}^t \times \left(\mathbf{\Psi_t} \circ \mathbf{{st}}^{k-1}\right) \times \mathbf{TR} & \label{eqn:SR}\\
	    & \mathbf{DT}^k_{ab} = \frac{1}{\sqrt{\mathbf{ST}^{k}_{aa}} \cdot \sqrt{\mathbf{ST}^{k}_{bb}}} \quad
	    \mathbf{DR}^k_{ab} = \frac{1}{\sqrt{\mathbf{SR}^{k}_{aa}} \cdot \sqrt{\mathbf{SR}^{k}_{bb}}} &
    \end{eqnarray}

In the above equations,  $\mathbf{\Psi}_r$ (resp., $\mathbf{\Psi}_t$) refers to a square matrix $n_r \times n_r$ (resp., $n_t \times n_t$) where all the elements are set equal to the mutual reinforcement factor $\psi$, with the exception of the diagonal, where the elements are set to 1; the symbol $\mathbf{TR}^t$ represents the transpose of $\mathbf{TR}$. Each step of the tag similarity computation can thus be performed by means of a simple matrix product. This result, coupled with the empirical observation that only a few iterative steps are required to reach convergence (Section~\ref{sub:scalability}), makes our similarity metric suitable in practical contexts. This conclusion is even more valid if we consider that, in our approach, tag similarity computations are performed offline.


\section{Evaluation}
\label{sec:evaluation}

In order to evaluate the performance of our approach, we built a prototype in Java and MySQL and we performed experiments using two well known social tagging websites, i.e., Bibsonomy and CiteULike. The experiments we carried out aimed at answering the following questions: {\em (i)} Is our approach  able to increase the accuracy of  searches? And, if so, to what extent does the improvement depend on the underlying similarity metric in use? {\em (ii)} Does our approach scale to large folksonomies?

After presenting the two datasets used for experimentation, we will address each of the above questions in turn.

\subsection{Datasets}
\label{sub:dataset}

We performed experiments on the following two datasets extracted from real large-scale folksonomies.

\begin{itemize}
\item {\em Bibsonomy} ({\small \tt http://www.bibsonomy.org/}) is a social \linebreak bookmarking website promoting the sharing of both scientific references and general URLs. We downloaded a snapshot of this Web site in June 2009, containing 648,924 bookmarks\footnote{In this context, a bookmark is defined as  a triplet $\langle u, r, tSet \rangle$, where $tSet$ is the set of tags originally assigned by the user $u$ to label the resource $r$.} and 4,696 users who had tagged 578,587 scientific references overall, using 147,076 distinct tags.

\item {\em CiteULike} ({\small \tt http://www.citeulike.org/}) is a social \linebreak bookmarking website that aims at promoting and developing the sharing of scientific references amongst researchers. We downloaded a snapshot dataset with 57,053 users, 1,928,302 papers, 401,620 different tags, and 2,281,609 bookmarks.

\end{itemize}

\subsection{Accuracy of User Searches}
\label{sub:userSerches}

\subsubsection{Simulation Setup}

The first experiment we conducted aimed at determining the ability of our approach to retrieve resources of relevance to the user querying the folksonomy. This experiment was enacted as follow: we  split each dataset into two different ones, called {\em train set} and {\em test set}; the split was performed multiple times at random, with the former containing 90\% of bookmarks, and the latter containing the remaining 10\%. After this split, we considered two different versions of the involved datasets, which we will refer to as {\em original} (un-modified) ones and  {\em enriched} ones. The enriched version $f^+$ of each dataset $f$ was obtained as follows. We computed similarities between all pairs of tags in the train set. After this, we examined all  bookmarks in the original  $f$; each bookmark $\langle u, r, tSet \rangle$ was then enriched by adding to $tSet$ the $k$ tags that our approach would recommend, as per Equation~\ref{eqn:Score}. In this experiment, $k$ was set equal to $\lceil 0.5 \cdot tSet \rceil$ if $tSet > 6$, equal to 3 otherwise.

In so doing, we simulate the automatic expansion of user-selected tags, as it would happen  when labeling resources. The folksonomy $f$ would thus be induced to grow to the enriched  $f^+$,  an emerging folksonomy containing a more common and widely accepted vocabulary.

Having performed this preparatory step, we considered $f$ and, for each bookmark $\langle u, r, tSet \rangle$ in the test set, we used $tSet$ to query the train set and retrieve the $q$ resources most relevant to the query. Note that, according to our approach, the query tag set $tSet$ was first expanded with the $k$ most related tags,  $k$ set as above. To determine the relevance of a resource to a query $tSet$, we computed the  {\em TF-IDF} coefficient assigned to such resource for each query tag  $t_j \in tSet$, then summed up these values. The $q$ most relevant resources were then offered to the user as answer to her query. We have experimented with   $q \in \{ 5, 10, 20 \}$, and  measured the percentage of times  the searched resource $r$ appeared in the top  $q$ retrieved resources. We call this measure {\em retrieved ratio}. We repeated the same procedure for the folksonomy $f^+$ to see if the corresponding retrieved ratio increased (or decreased) with respect to the one of $f$.

The above process follows the intuition that, if a user {\em labeled} a specific resource $r$ with the set of tags $tSet$, then $tSet$ would very likely be the set of tags such user would employ to {\em query}  the folksonomy when willing to retrieve $r$. However, due to the number of synonyms, homonyms and polysemies, as well as the heterogeneity of users, $r$ could have been described by users with different tags. This implies that $tSet$ could be unsuitable to retrieve $r$ per se, and thus tag expansion, performed both when labeling and when querying the system (i.e., $f^+$), should yield better results (i.e., increased retrieved ratio).

We repeated the described process 10 times over different train and test random splits of the datasets. The results we describe next represent the average values of these runs.

\subsubsection{Results}

\begin{figure}%
    \centering
    \includegraphics[width=.50\columnwidth]{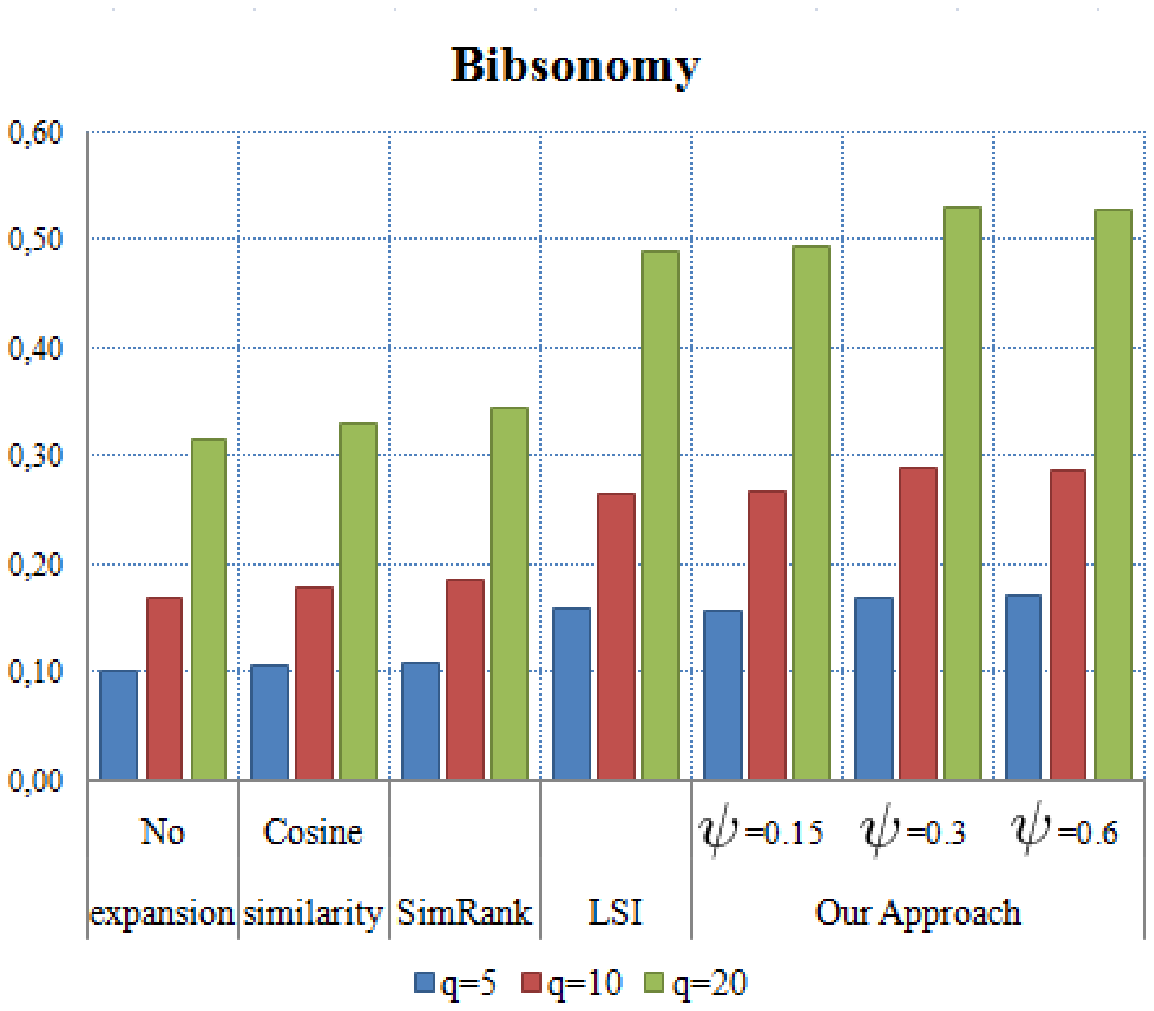}%
    \includegraphics[width=.50\columnwidth]{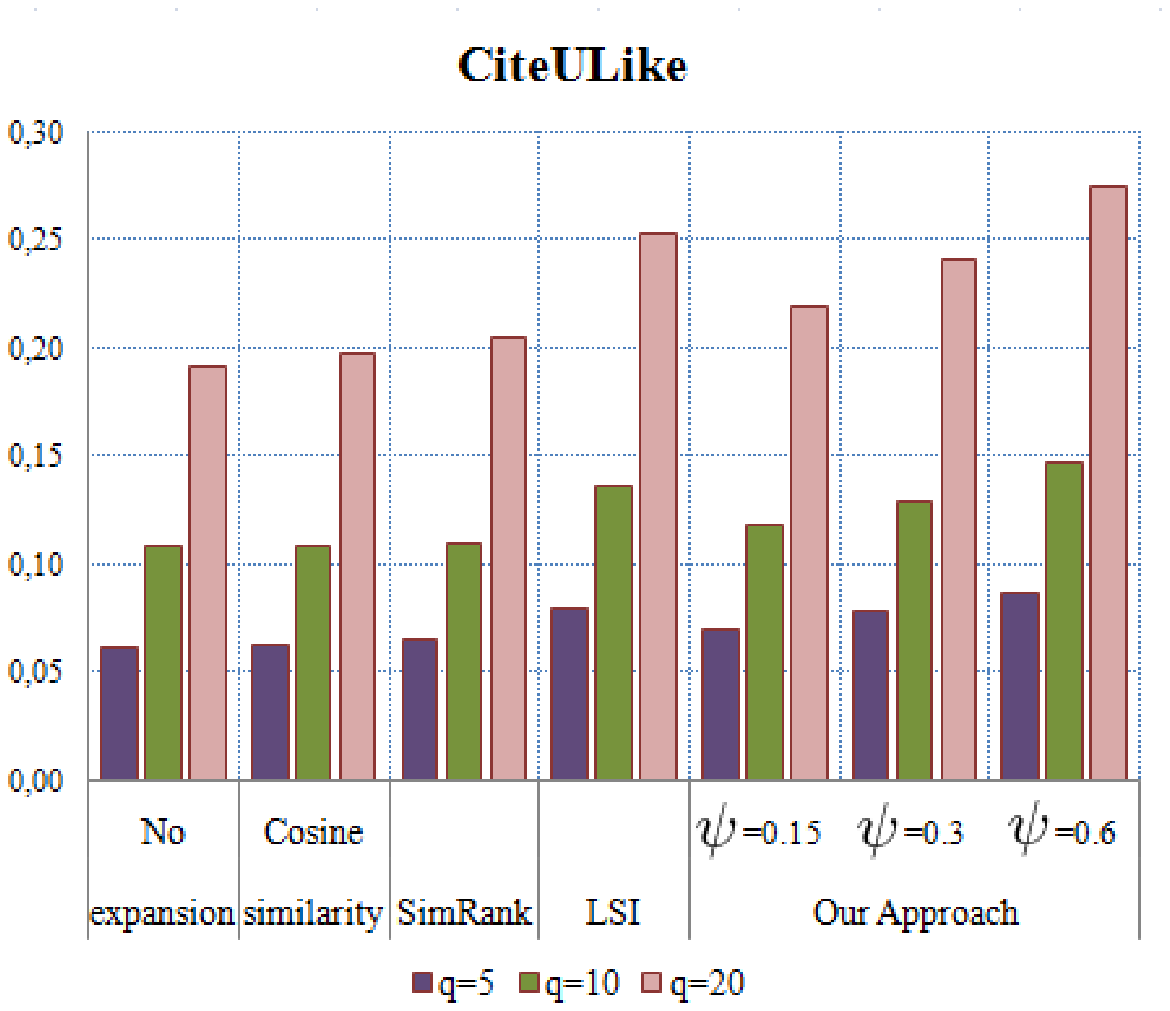}%
    \caption{Retrieved Ratio on Bibsonomy and CiteULike}%
    \label{fig:retrievedRatio}%
\end{figure}

Figure~\ref{fig:retrievedRatio} shows how the retrieved ratio of our approach varies across the two datasets, for different values of $\psi$,  with respect to the case where no enrichment was performed. The same figure also illustrates a comparison of our approach with respect to cases where cosine similarity, Latent Semantic Indexing, and SimRank were used as the underlying tag similarity metric. The following two main observations can be drawn:

\begin{itemize}

\item The lowest retrieved ratio is the one obtained when no enrichment is performed. In particular, the retrieved ratio is up to 70\% better when using an enriched folksonomy than when used an unmodified one. This means that, by transparently enriching the folksonomy when labeling resources, we successfully induce the construction of a denser and more meaningful  folksonomy, over which resource retrieval performs better overall. This result underlines the  importance of supporting users in their tagging and querying activity.

\item Within the enriched folksonomy, the approach based on our novel tag similarity metric outperformed all others, followed by LSI, while SimRank and cosine lag behind. More precisely, results obtained by applying our similarity metric are up to 50\% better than those obtained by applying cosine similarity or SimRank, and up to 8\% better than those obtained by applying LSI.

\end{itemize}

\subsection{Scalability}
\label{sub:scalability}

\subsubsection{Simulation Setup}

As previously pointed out, the highest cost caused by our approach lies in the computation of pairwise tag similarities.  In Section~\ref{sub:Phase1}, we have shown that our metric is recursive and that each step is no more expensive than the computation of classical similarity measures, such as cosine similarity. We have also proved that our formulation is convergent (Theorem~\ref{the:conv}). However, it is necessary to investigate how many steps are necessary to reach  convergence in practice.

To experimentally perform this computation, we have defined the following parameters
    \begin{equation}
    	\delta_t^k = \frac{ || \mathbf{{st}}^{k} - \mathbf{{st}}^{k-1} ||_1 }
        { || \mathbf{{st}}^{k} ||_1 } \quad \mbox{and} \quad
        \delta_r^k = \frac{ || \mathbf{{sr}}^{k} - \mathbf{{sr}}^{k-1} ||_1 }
        { || \mathbf{{sr}}^{k} ||_1 }
    \end{equation}

Here, $\mathbf{{st}}^{k}$ (resp., $\mathbf{{sr}}^{k}$) are the tag-tag (resp., resource-resource) similarity matrices at the $k^{th}$ step (see Equation~\ref{eqn:matrices}), whereas symbol $|| \cdot ||_1$ indicates the \mbox{1-norm} of a matrix.

\subsubsection{Results}

\begin{figure}%
    \centering
    \includegraphics[clip=true, trim =90 0 95 0, width=.50\columnwidth]{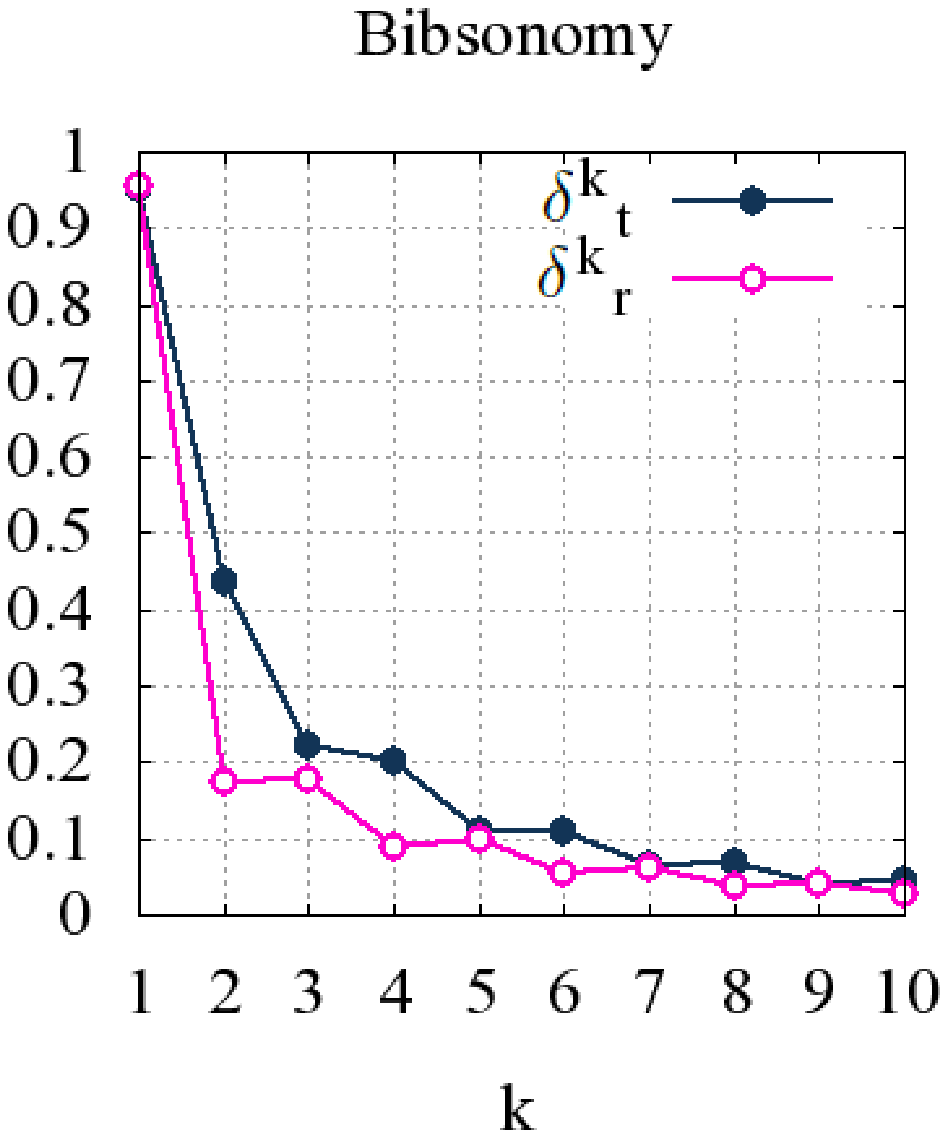}%
    \includegraphics[clip=true, trim =90 0 95 0, width=.50\columnwidth]{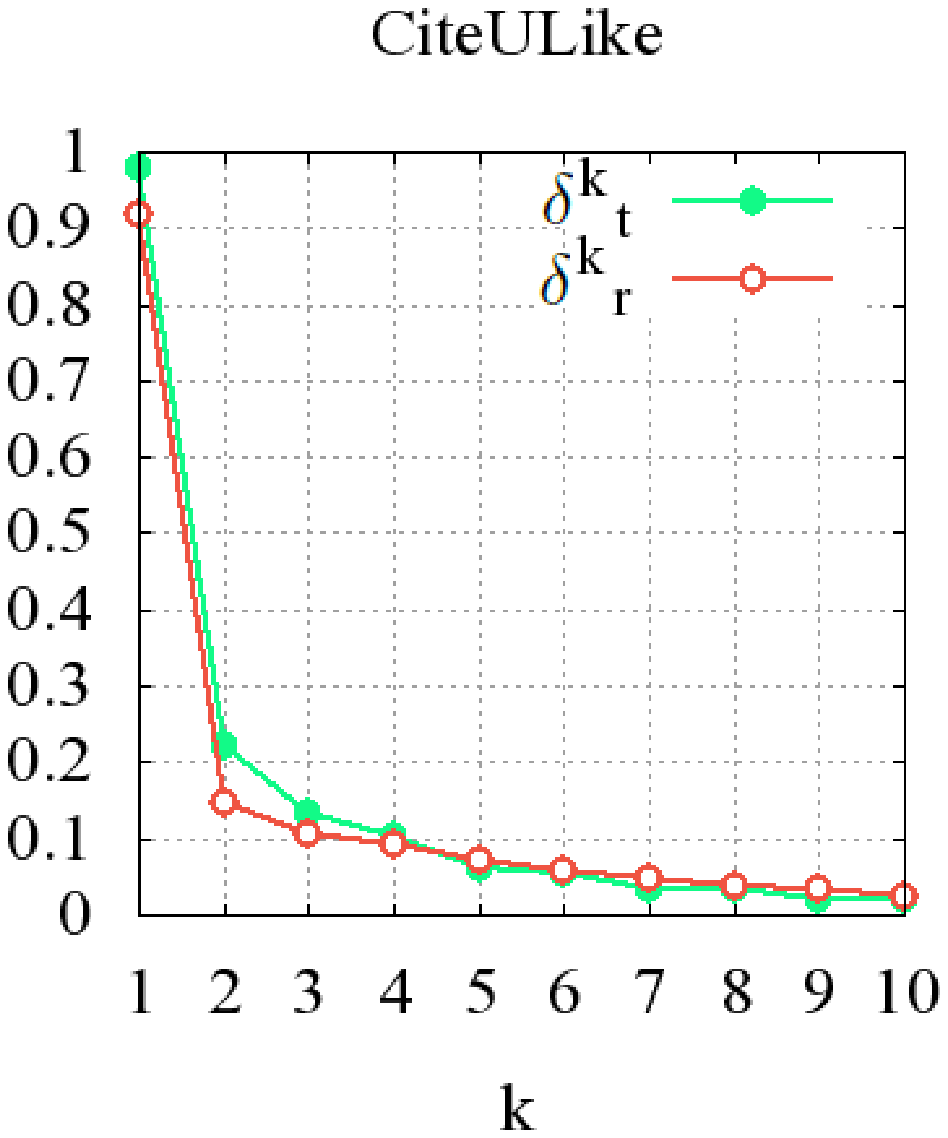}%
    \caption{Scalability of our Approach on Bibsonomy and CiteULike}%
    \label{fig:residual}%
\end{figure}

Figure~\ref{fig:residual}  plots the variation of $\delta_t^k$ and $\delta_r^k$ as $k$ increases, for each of our datasets. As shown, in the practical settings we have experimented with, the computation of our similarity metric exhibits very fast convergence. As an example, across all  considered datasets, $\delta_t^k$ and $\delta_r^k$ are less  than 0.1 after just six iterations. In other words, in  the datasets we used, by accepting a negligible error in the similarity computation, we can stop our iterative procedure in less than six iterations. Using a server equipped with a quad-core processor and 32GB of RAM (which is much smaller than any server deployed in practice by actual businesses), we computed all similarity measures across all two datasets in less than 48 hours. As the computation of pairwise tag similarity is performed periodically (e.g., weekly) offline, this result confirms that our similarity measure is scalable and well suited to be applied even when operating in large folksonomies.


\section{Conclusions}
\label{sec:conclusions}

In this paper, we have proposed an approach that enables the effective retrieval of resources within folksonomies. The approach relies on an innovative tag similarity metric that is based on the mutual reinforcement principle. This metric is used both when users label resources, so to automatically enrich the user-selected tag set with highly-related tags already present in the folksonomy, and when users query the folksonomy. Our experimental evaluation has demonstrated that the accuracy of searches entailed by our metric are neatly higher than those achieved by applying classical metrics, thus confirming its suitability in scenarios characterized by power-law distributions of tags (as is the case in many real world folksonomies). Finally, the computational cost of our iterative approach is limited, as convergence is guaranteed, and  in practice reached after a handful of iterations.

\bibliography{short-sigproc}
\bibliographystyle{plain}

\end{document}